\documentstyle[prl,aps,multicol,epsf]{revtex}
\begin{document}
\draft
\title{Single Carbon Nanotube--Superconductor 
Entangler:\\ noise correlations and EPR states.}
\author{V. Bouchiat$^{a}$, N. Chtchelkatchev$^{b,c}$, D.
Feinberg$^{d}$, G.B.  Lesovik$^{b,c}$, T. Martin$^{c}$,
and J. Torr\`es$^{c,e}$,} 
\address{$^{a}$Centre de
Recherches sur les Tr\`{e}s Basses Temp\'eratures, CNRS, BP166X,
38042 Grenoble, France}
\address{$^{b}$L.D.\ Landau Institute for Theoretical Physics,
Russian Academy of Sciences, Kosygina Str.\ 2, 117940, Moscow, Russia.}
\address{$^c$Centre de Physique Th\'eorique et Universit\'e de
la M\'editerran\'ee, Case 907, 13288 Marseille, France}
\address{$^{d}$Laboratoire d'Etudes des Propri\'et\'es
Electroniques des Solides$^{*}$ , CNRS, BP166, 38042 Grenoble,
France}
\address{$^e$Departement de Physique, Universit\'e de Sherbrooke,
Sherbrooke QC J1K2R1, Canada}  
\date{\today}
\maketitle

\begin{abstract}
{We propose a device which implements a solid-state nanostructured electron entangler.
It consists of a single-walled carbon nanotube connected at
both end to normal state electrodes and  coupled in its middle
part to a superconducting nanowire. Such a device  acts as an
electronic beam splitter for correlated electrons originating
from the superconductor.  We first show that it can be used to
detect positive (bosonic--like) noise correlations in an
fermionic system. Furthermore, it provides a source for entangled
electrons in the two arms of the splitter. For generating
entangled electron states, we propose two kinds of setup based
either on spin or energy filters. It respectively
consists of ferromagnetic pads and of a system of electrostatic
gates which define quantum dots.  The fabrication of this device
would require state-of-the-art nanofabrication techniques,
carbon nanotube synthesis and integration, as well as atomic
force microscopy imaging and manipulation.}  
\end{abstract} 
\begin{multicols}{2}  

\section{Introduction}

In the last decade, photon
entanglement has triggered the proposition of new information processing schemes
based on quantum mechanics \cite{revue_academic}. Indeed emerging fields such as quantum
cryptography and quantum communication \cite{crypt,telep} are based on
particle entanglement. On the other hand, concrete proposals for
quantum computing, based on electron transport and electron interactions in condensed
matter have been recently presented.
\cite{recher_sukhorukov_loss,recher_loss,lesovik_martin_blatter}.
Among them, devices taking advantage of the macroscopically
coherent wave of superconductivity \cite{revue_shnirman} are
promising  candidates for the practical realization of a fully
solid-state quantum bit  \cite{Qbit-JJ}.   

Recently, the proposal of Ref. \cite{lesovik_martin_blatter}
showed that the generation mechanism for entangled electrons
pairs emerging from a superconductor could provide a rather
robust alternative to photon entanglement. 
It was first shown \cite{Torres} 
that a normal metal fork attached to a superconductor 
can exhibit positive correlations. Positive correlations had
been attributed to photonic systems in the seminal Hanbury-Brown
and Twiss experiment \cite{HBT}. In our case, positive correlations
 arise because evanescent Cooper
pairs can be emitted on the normal side, due to the proximity
effect \cite{proximity}. These Cooper pairs can either decay in
one given lead, which gives a negative contribution to noise
correlations, or may split at the junction on the normal
side with its two constituent electrons propagating in different
leads. This latter effect constitutes the justification for
positive noise correlations. Moreover such a mechanism
generates entangled and delocalized electron 
pairs\cite{lesovik_martin_blatter}. Two
electrons originating from a broken Cooper pair bear entangled energy
and spin degrees of freedom. This provides a solid state analog of 
Einstein-Podolsky-Rosen (EPR) states which were proposed to    
demonstrate the non-local nature of quantum mechanics \cite{EPR}.
Both theoretical proposals  -- for positive correlations and
for EPR entanglement in electronic systems -- are in need for
 experimental observation. The purpose of the present paper is to
define a solid-state device together with a detection setup of these both quantum effects.
It is based on a Single Walled
Carbon Nanotube (SWNT) coupled to a Superconducting electrode 
(nicknamed in the following the S-SWNT device). 

Troughout the paper a single electron description of transport
will be adopted. This choise is motivated in several ways. 
First, single electron 
scattering theories have been quite successful so far in 
describing the transport properties of carbon nanotubes. 
An example is the seminal Fabry-Perot experiment of 
Ref. \cite{nature_nt_fabry_perot} which can be interpreted 
with a ballistic propagation picture. Second, we intend
here to describe a rather complex device: electrons are injected  
from a superconductor in a nanotube, possibly with additional scattering 
elements or filters. It therefore makes more sense to first 
enquire what the 
transport properties are from the scattering theory point of view, 
rather than to go immediately to a correlated electron description.
Nevertheless, evidence of Luttinger liquid behavior in tunneling
geometries involving
metallic SWNTs has been proposed \cite{kane_balents_fisher}.
The experimental measurement of the tunneling $I(V)$
characteristics \cite{mceuen} yields density of 
states exponents which are consistent with
strong correlations. Below we address briefly
when the single electron picture is expected to hold,
and what features of Luttinger liquid theory could
affect the S-SWNT device.

\begin{figure} 
\centerline{\epsfxsize=6.0cm \epsfbox{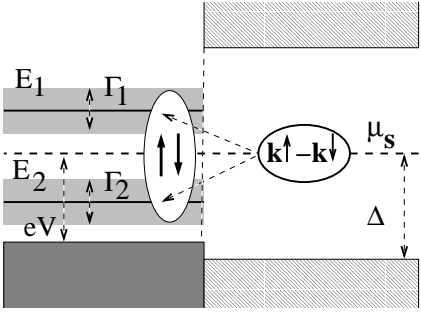}}  
\narrowtext\vspace{2mm} 
\caption{Transfer of a Cooper pair on two quantum energy levels
$E_{1,2}$ with a finite width 
$\Gamma_{1,2}$. The spacing between the 
two energies is assumed to be well within the gap
to avoid quasiparticle excitations. The transfer
of a Cooper pair gives an entangled state in the dots.
The source drain voltage $eV$ for measuring noise 
correlations is indicated.}   \end{figure}  

\section{Entanglement of electrons tunneling from a superconductor}
\label{entanglement_section}

A perturbative argument supports the claim 
that two electrons originating from the same Cooper pair are
entangled. Consider two quantum dots
(Energies $E_{1,2}$) next to a superconductor. The 
state of the latter is specified by the BCS wave function
$|\Psi_{BCS}\rangle=\prod_k (u_k+v_kc^\dagger_{k\uparrow}  
c^\dagger_{-k\downarrow})|0\rangle$. Tunneling to the dots
is described by a single electron hopping Hamiltonian:
\begin{equation}
H_t=\sum_{k\sigma}[t_{1k}c^\dagger_{1\sigma}+
 t_{2k}c^\dagger_{2\sigma}] c_{k\sigma}+ h.c.~,
\end{equation} 
 with $c^\dagger_{k\sigma}$ creates an electron with spin $\sigma$.
Now let assume that the transfer Hamiltonian acts on a single
Cooper pair. Using the T-matrix to lowest ($2$nd) order, the
wave function  contribution of the two particle state with one
electron  in each dot reads:
\begin{eqnarray}
|\delta \Psi_{12}\rangle
&=&
\sum_k 
v_{k}u_{k}t_{1k}t_{2k}
\left({1\over i\eta
-E_{k}-E_1}+
{1 \over i\eta-E_{k}-E_2}
\right)\nonumber\\
&~&~~\times 
[c^\dagger_{1\uparrow}c^\dagger_{2\downarrow}-
c^\dagger_{1\downarrow}c^\dagger_{2\uparrow}] 
|\Psi_{BCS}\rangle
\label{entangled_dots}
\end{eqnarray}
where,
$E_k$ is the
energy of a Bogoliubov quasiparticle.  The state of Eq.
(\ref{entangled_dots}) has entangled spin degrees of freedom,
a result of the spin symmetry of the
tunneling Hamiltonian. From the superconductor,
$H_t$ can only produce singlet states in the dots.  
We now present a device where this entangled state propagates
along metallic wires.

\section{Noise correlations in the S--SWNT device}
\label{noise_section}

\begin{figure} 
\centerline{\epsfxsize=8.0cm \epsfbox{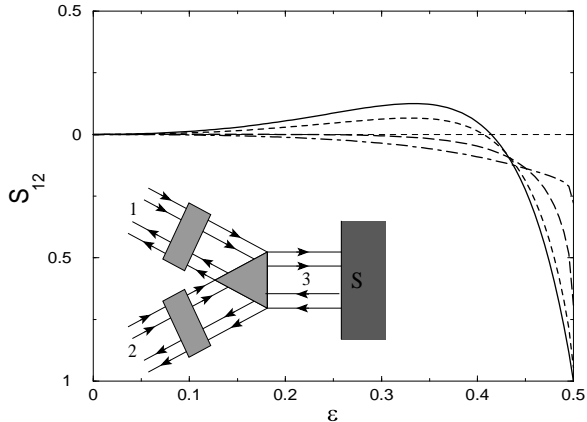}}  
\narrowtext\vspace{2mm} 
\caption{\label{fig2}Dependence of the current--current
noise correlations between SWNT leads in presence of channel
mixing, as a function of the transparency of the beam splitter
($\epsilon=0$ corresponds to a totally opaque splitter):
$\delta=0$ (solid line), $\delta=0.1$ (short dashed line), 
$\delta=0.2$ (long dashed line), $\delta=0.3$ (dashed--dotted
line).  Inset: schematics of the two channel beam splitter}  
\end{figure}

Consider a system consisting of a superconducting electrode which is
contacted locally to the middle of a single-walled carbon nanotube.
(SWNT). Current can in principle  be measured at each extremity of the
1D conductor. From the point of
view of the Landauer--Buttiker scattering approach to quantum
transport  \cite{landauer-buttiker}, it
consists of scattering elements, including a ``beam splitter'',
and a normal superconductor interface (\ref{fig2}).
A more realistic drawing of this setup is also
depicted in Fig. \ref{fig3}. 
Note that this device is analogous to the one 
used in the fermion analog of the Hanbury--Brown
and Twiss experiment \cite{martin_landauer,HBT_fermion}, except
that the electron source is a superconductor.

The scattering matrix which specifies the amplitudes 
of the incoming and outgoing states at the junction
does not provide any information about entanglement, 
because of its single-electron character. Yet we argue 
here (and below) that entanglement is implicit in
correlations measurements.
The generation of entangled, non-local electronic states  
requires that a substantial fraction of Cooper pairs distribute 
their electrons in the two leads, rather than in the
same one. A suggested diagnosis of the presence
of Cooper pairs in the two arms of a nanotube
lies in the noise correlator:
\begin{equation}
S_{12}(\omega)
=
\int_{-\infty}^{+\infty} dt e^{-i \omega t}
\Bigl(
\langle I_1(t) I_2(0) \rangle
- \langle I_1 \rangle \langle I_2 \rangle
\Bigr)
~,
\label{eq_noise_def}
\end{equation}
evaluated at zero frequency.
Here, $I_i(t)$ denotes the current operator 
in lead $i$.
The positive noise correlations predicted
in a single channel NS junction\cite{Torres}
constitute a direct consequence of these tunneling 
processes. For the SWNT beam splitter, 
the presence of two propagating channels
at the Fermi energy in each lead 
requires that we address the issue of channel
mixing due to impurities/geometrical scattering.
The addition of several 
transverse channels \cite{Gramespacher} 
may destroy the 
positive noise correlations 
which signal the presence of
entangled electron pairs.

Because of the two channels in each lead, a $4\times4$
scattering  matrix fully characterizes this branched 
N-S junction (no spin-flip scattering). 
Denote by
$s_{ij\alpha\beta p p'}$ the scattering amplitude for a particle
$p$ ($p=e,h$, electron or hole) incident from channel $\beta$
associated with lead $j$, transferred in channel $\alpha$ of
lead $i$ as a particle  of type $p$. Using the scattering
formulation of quantum transport together with the
Bogoliubov--de Gennes  transformation
\cite{torres_martin_lesovik}, 
the zero frequency noise correlations below the gap
become:  
\begin{eqnarray}
&&S_{12}(0)=  \frac{2e^2}{h}\int_0^{eV} dE
\sum_{i,j=1,2}\sum_{\alpha_1,\alpha_2}\\
&& \times\left( s^*_{1 i \alpha_1 \alpha_i ee} s_{1 j \alpha_1
\alpha_j eh} - s^*_{1 i \alpha_1 \alpha_i he} s_{1 j \alpha_1
\alpha_j hh} \right) \\ && 
\times\left( s^*_{2 j \alpha_2 \alpha_j eh} 
s_{2 i \alpha_2 \alpha_i ee} - s^*_{2 j \alpha_2 \alpha_j hh}
s_{2 i \alpha_2 \alpha_i he} \right) 
\label{noise correlation nanotube}
\end{eqnarray}
Our model device is depicted in the 
inset of Fig. 3. It is composed of four scattering 
elements. Perfect Andreev reflection  occurs at the 
N-S interface (with no channel mixing), and two 
independent beam splitters (for electrons and holes)
describe the
connection  to the SWNT. 
The splitters transmission to the two leads
is controlled by a single parameter $\epsilon$ 
($\epsilon=0.5$ for maximal transmission) as in
Ref. \cite{Torres}.  
Channel mixing is then included
within each tube using a numerical random matrix scheme 
\cite{Siffert}, here represented by the two rectangles
on the left hand side of Fig. 3.  
After specifying each S--matrix of the subsystems, 
those are combined to yield a $4\times 4$
S--matrix which characterizes the propagation of electrons
and holes from $1$ to $2$.

The scheme for including mode mixing and 
backscattering in a $2$ mode quantum wire 
generates numerically random (unitary) 
scattering matrices:
${\bf s}= \exp(i {\bf h})$,
where ${\bf h}$ is  a random, Hermitian matrix.
It only makes sense to compare different
samples with the same amount of disorder. A perfect,
$N$-channel conductor can be described by a scattering 
matrix ${\bf s_{0}}$ which is a block matrix with identity 
matrices on the off diagonal blocks, zero otherwise. 
The corresponding matrix ${\bf h_{0}}=-i\ln{\bf s_{0}}$
can be found, but is not unique due to
the periodicity  of the exponential function. 
We generate unitary, random matrices by adding a small
random perturbation to ${\bf h_{0}}$:
${\bf s}=\exp i[{\bf h_0}+\delta\pi\sqrt{2}{\bf
R}]$,
where ${\bf R}$ is a random hermitian 
matrix with a unit norm. The
limitation  $\delta<1/\sqrt{2}$ then insures that 
the S--matrix generated in this manner 
has an increased amount of disorder 
starting from $\delta=0$.
In this manner, samples with
a specific conductance, or equivalently, 
a specific mean free path per sample length,
are generated.

The zero frequency noise correlations are obtained using 
Eq. \ref{noise correlation nanotube} and 
averaging over $200$ sample configurations. 
The results are illustrated by the curves
of Fig. \ref{fig2}, which show the averaged 
zero frequency noise correlations as a function 
of the parameter $\epsilon$ which parameterizes 
the connection of the beam splitter to the
two nanotube arms. The different curves correspond
to varying degrees of channel mixing.   

When an infinitesimal mixing or backscattering $\delta$ is included in each
nanotube arm (weak disorder), 
positive noise correlations   are observed
as in  Ref. \cite{Torres} for a vast
majority of the coupling parameter $\epsilon$. 
For non zero $\delta$, the disorder in the
nanotube arms is increased, and  it restricts 
the possibility for
positive noise correlations. The positive correlations are mostly
reduced in amplitude, while occurring for the same ranges of
$\epsilon$. The  range of $\epsilon$ with
positive correlations  is  reduced, and eventually vanishes
completely  for stronger mixing. Note that the disorder parameter
$\delta=0.3$ lies in the strong scattering regime.
Nevertheless, we conclude
that this system is sufficiently robust because a weak amount of
disorder does not entirely spoils the  effect.

\section{The S-SWNT device: a Cooper pair splitter}
\label{splitter_section}

\begin{figure} 
\centerline{\epsfxsize=8.0cm \epsfbox{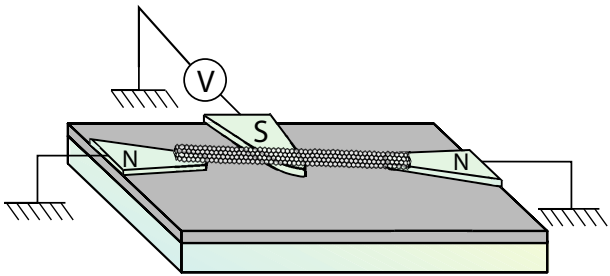}}  
\narrowtext\vspace{2mm} 
\caption{\label{fig3} Schematics of the S-SWNT device: the
nanotube is deposited on top of a superconducting ``finger'',
and is connected to metallic leads.  The measurement of noise
correlations implies that a bias is imposed between 
the superconducor and the normal metal electrodes}  
\end{figure}

The practical implementation of this device 
\cite{Torres} consists of a superconducting reservoir
connected to two normal metal leads (see Fig. \ref{fig3}). 
Ballistic propagation in these normal leads
is optimal for the detection of electron 
entanglement via a noise correlation measurement. The
 Single Walled Carbon Nanotube (SWNT) appears to be a good candidate
for implementing the normal lead. Indeed SWNTs can be considered 
as quasi-ideal one dimensional electron
waveguides\cite{Bezryadin,Bachtold_PRL,nature_nt_fabry_perot},
with either metallic or semiconductor behavior depending on
their helicity.  Metallic SWNTs have two propagating modes with
equal velocity  at the Fermi level and can exhibit 
quasi-ballistic transport \cite{White_Todorov}.
On the other hand semiconducting SWNTs can be
electrostatically \cite{transistor_tans,NT_NASA} or chemically doped \cite{Dai_dope}
to make a single channel conductor, which is however 
more sensitive to disorder. Nanotubes also present the advantage that they
 can be reliably assembled into complex integrated circuits 
\cite{Bachtold_NTlogic,avouris_device} and withstand
 controlled intra-molecular functionalization 
\cite{Dai_dope,SETintraNT}. These recent advances were
realized   using  state-of-the art nanofabrication techniques:
electron beam lithography, together with the alignment,
sensitivity and manipulation abilities provided by Atomic
Force Microscopy (AFM). Furthermore, SWNTs
can be connected {\it in situ} during their synthesis by letting 
them grow from superconducting electrodes. Such process
 involves a Chemical Vapor Deposition route \cite{marty}. The
remaining problem is to create two SWNT--superconductor contacts
 in order to achieve an ``N-S forked''-shaped device. There
have already experimental evidence that such contacts can be made with  a
sufficiently high transparency so that a supercurrent can flow
through the device due to the proximity effect
\cite{proximity supra-NT}.  There are two fundamental requirements 
for this connection. First,  the  two constituent electrons of
a Cooper pair should be  transmitted symmetrically in the  two
leads. Second, the distance between  these leads needs to be
smaller than the size of  a Cooper pair. The fabrication of a 
connection of two SWNTs with comparable  insulating barrier
thicknesses  at the surface of the superconductor,
remains challenging \cite{nature_nt_fabry_perot}. In fact,  
the connection to two separate 
SWNTs at a superconductor surface 
can be avoided altogether by using only one SWNT coupled 
to the superconductor
in its middle part. 
  
Either it could be bent on the 
superconductor interface (and the radius of 
curvature of the bent is small or comparable 
to the superconductor coherence length), or 
the nanotube is kept straight while a 
superconducting finger is placed in contact on top.
In the latter case, the width of
the finger is chosen to be  comparable  to the coherence
length (Fig. \ref{fig2}) so that electrons/holes injected
from one normal contact can be Andreev reflected into
holes/electrons in the opposite normal contact.

\begin{figure} 
\centerline{\epsfxsize=6.0cm \epsfbox{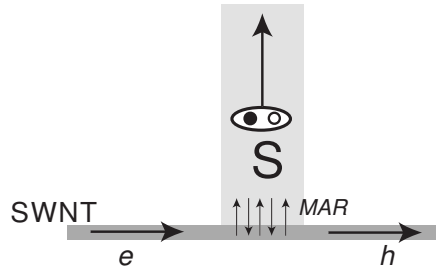}}  
\narrowtext\vspace{2mm} 
\caption{\label{fig-MAR} Andreev reflection process
which involves the two extremities of a nanotube deposited 
on top of a superconductor: the incoming electron (left) 
suffers multiple Andreev reflections in the region which is
contacted with the superconductor, and exists as a hole on the
other side.}
\end{figure}

Several proposals involving the connection of two nanotubes
to a superconductor have appeared recently
in the literature \cite{recher_loss,bena}, which aim at
generating entangled states in the two separated nanotubes.
Such a proposal suffers from the fact that the Andreev
amplitude -- or equivalently the Cooper pair emission
amplitude -- is strongly reduced by geometrical factors
\cite{recher_loss,choi_bruder_loss}. Typically, such factors scale
like $(k_f d)^{-2}$, with $d$ the distance between the two
nanotube--superconductor contacts and $k_f$ the Fermi wavevector in the
superconductor. Due to this geometrical
factor, it is necessary for the two nanotubes to be a few nanometers
apart. Here, it is argued that first, the manipulation of a single
SWNT contacted on the superconductor is likely to be easier
to handle experimentally. Second, and more importantly, the
deposition  of a quasi one--dimensional object in contact with  a
superconductor, does not suffer from geometrical factors.  In fact, an
incident electron which arrives in
the region where the nanotube is in contact with the
superconductor may suffer multiple Andreev reflections (see Fig. \ref{fig-MAR}), provided that the
interface transparency is good enough.
Consequently, an Andreev reflected hole will be emitted
at the other extremity of the tube, even though $k_f d >> 1$. This
mechanism is somewhat related to the idea of a proximity effect in the
nanotube \cite{recher_loss}, but its description requires further
theoretical study.

Note that at the two locations (right and left on 
Fig. \ref{fig-MAR})
where the tube meets the superconductor, normal scattering is
likely to occur  because bending the nanotubes generates
topological  defects in the latter \cite{Bezryadin,SETintraNT}. 
Paradoxically, this
scattering is necessary in order to break translational
invariance, so that the electron incoming from the left
can indeed be ``Andreev-transmitted'' into a hole on the right
(an not be Andreev-reflected on the left). 
Alternatively, the superconducting material could be 
evaporated on top of the nanotube. 

\section{Entanglement diagnosis and scattering theory}
\label{filters_bell_section}

\subsection{``Wavepacket reduction'' with selective filters}
\label{filters_subsection}  

The positive correlations of section \ref{noise_section} were 
due to the splitting of a Cooper pair,
with the two constituent electrons being redistributed among the
two arms of the nanotube. Entanglement will occur if a Cooper 
pair is prohibited to enter a given lead as a whole. 
In Fig. 5, two specific setups which 
implement a selective filtering 
of the correlated electrons are proposed: 
a) using two ferromagnetic metal pads (with magnetizations
in opposite directions) in each lead (Fig. 5a) which effectively
block propagation of electrons bearing the opposite spin, b) exploiting the
difference in kinetic energies of electron and hole
quasiparticles and thus positioning either a single
 gate controlling the energy of intramolecular quantum dots 
which have been defined either by 
intramolecular scatterers \cite{QD_nanotube,defect_NT}
or by defining set of gates on the leads (Fig. 5b) that will 
act as a Fabry-Perot-like energy filters. 
\begin{figure} 
\centerline{\epsfysize=6.0cm \epsfbox{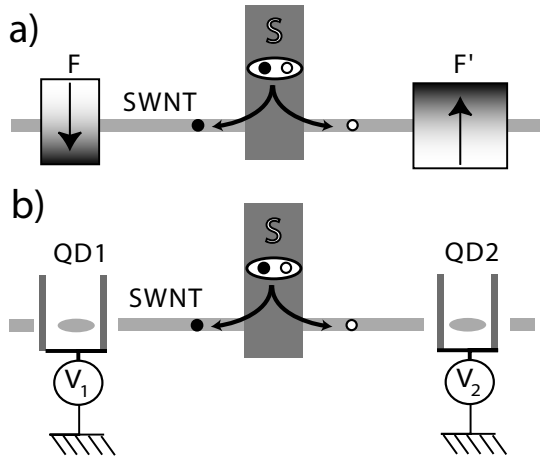}}  
\narrowtext\vspace{4mm} 
\caption{Schematics of two complementary detection
 setups for measurement of entangled electrons in the S-SWNT device: a) with spin 
filtering,implemented by ferromagnetic pads having different sizes denoted by F1 and F2 ; b) with
Fabry Perot filters,implemented by quantum dots denoted by QD1 and QD2 where the gate voltages V1 and V2 on each side
selects quasiparticles (say, left, energy $E$) and  quasi hole
(right, $-E$)} 
\end{figure}
The extension of the setup proposed in Ref.
\cite{lesovik_martin_blatter} to two--channel leads is 
straightforward.
With filters, this forked superconductor geometry is
a two terminal one, where electrons with a given
spin from lead $1$ are converted into holes with an opposite
spin in lead $2$.  The noise correlations between normal leads
$1$ and $2$ {\it exactly} corresponds to the noise in {\it one}
of the leads: 
\begin{eqnarray} S_{12}(0)
&&= {8e^3\over h}
{\mathrm Tr} 
\left[
{\bf s}_{\alpha,\alpha^\prime}{\bf
s}^\dagger_{\alpha,\alpha^\prime}
( 1-
{\bf s}_{\alpha,\alpha^\prime}{\bf
s}^\dagger_{\alpha,\alpha^\prime} )
\right]V~,
\nonumber\\
&&= {8e^3\over h}
\sum_{\gamma=1,2} T_\gamma(1-T_\gamma)V=S_{11}(0)~.
\label{noise_one_channel} 
\end{eqnarray} 
with $V$ the applied bias. $T_\gamma$ are
``transmission eigenvalues'' of the $2\times
2$ matrix  ${\bf s}_{\alpha,\alpha^\prime}{\bf
s}^\dagger_{\alpha,\alpha^\prime}$, which correspond to the 
Andreev reflection probabilities in the so-called
eigen-channel representation \cite{martin_landauer}.
For ferromagnetic filters (SN-FF) with the
spin in F$_{1(2)}$ pointing up (down) $\alpha = \{e(h) \uparrow
1\}$ and $\alpha^\prime = \{h(e) \downarrow 2\}$ (the
propagation of other states is blocked). For the setup selecting
quasiparticles  and quasi--holes in leads N$_1$ and N$_2$ via
Fabry-Perot type filters, we have to sum over spins with $\alpha
= \{e \uparrow\!(\downarrow)\, 1\}$ and $\alpha^\prime = \{h
\downarrow\!(\uparrow)\, 2\}$. The decomposition of Eq.
(\ref{noise_one_channel}) leading 
to the scattering matrix eigenvalues 
has been exploited recently in the analysis 
of multiple Andreev 
reflection phenomenon in atomic point contacts
created with break junctions, thus uncovering the 
``mesoscopic code'' of such devices \cite{cron}.
The spin current which flows in
one branch is thus perfectly  correlated to the opposite spin 
current in the other lead \cite{feinberg}: $\langle\langle
(I_{\sigma 1}- I_{-\sigma 2})^2\rangle\rangle=0$.

The wave function which
describes entangled states in this two channel, two lead device
is now written for the two types of filters, in the
eigen-channel representation \cite{martin_landauer}.  For spin
filters, \begin{eqnarray}
|\Phi^{\rm spin}_{\varepsilon,\sigma}\rangle =
\sum_{\gamma=1,2}
&&\alpha_{\gamma}|\gamma;\varepsilon,\sigma;
-\varepsilon,-\sigma \rangle
\nonumber\\
&&+\beta_{\gamma} |\gamma;- \varepsilon,\sigma; 
\varepsilon, - \sigma \rangle~,
\label{wave_function_spin}
\end{eqnarray}
where the first (second) argument in $|\phi_1;\phi_2\rangle$
refers to the quasi-particle state in lead 1 (2) evaluated behind
the filters; $\gamma$ is the SWNT
eigen-channel index and $\sigma$ is a spin index; the
coefficients  $\alpha_{\gamma}$ and
$\beta_{\gamma}$ can be tuned by external parameters, 
e.g., a magnetic field.  Note that by projecting the spin
degrees of freedom  in each lead, the spin entanglement is
destroyed.  Nevertheless, the energy degrees of freedom are 
still entangled, and could in principle lead to 
a measurement of quantum mechanical non locality: a measurement
of energy $\varepsilon$ in lead $1$ projects the wave function 
so that the energy $-\varepsilon$ has to occur in lead $2$.
Such a measurement could be made connecting the device to 
a set of quantum dots/energy filters. However a direct
 analogy with Bell--type inequalities for photons with
crossed polarizers is not possible here. 

On the other hand, the energy filters do  
preserve spin entanglement:
\begin{eqnarray}
|\Phi^{\rm energy}_{\varepsilon,\sigma}\rangle =
\sum_{\gamma=1,2}&&
\alpha_{\gamma}|\gamma;\varepsilon,\sigma;
-\varepsilon,-\sigma \rangle
\nonumber\\
&&+\beta_{\gamma} |\gamma; \varepsilon, -
\sigma; - \varepsilon,\sigma \rangle,
\label{wave_function_energy}
\end{eqnarray}
Note that in principle, the electrons emanating from the
energy filters could be analyzed in a similar manner as in Bell
type measurements, using now spin filters with variable magnetization
orientation as a detection setup. the efficiency of spin filtering 
by connecting a ferromagnetic electrode on a SWNT has already been 
pointed out \cite{spin_SWNT}, while transport
of spin-polarized electrons in a carbon nanotube 
has been already experimentally observed\cite{alphenaar}.

\subsection{Entanglement detection: Bell inequalities}
\label{bell_subsection}

Strictly speaking, the measurement of perfect noise correlations
as illustrated in Eq. (\ref{noise_one_channel}) 
only constitutes a proof of charge correlation between the two
leads.  We now address the correlations of charge and spin.
In photon experiments, entanglement is detected in Bell setups,
where coincidence  measurements count the scattered photons along
different polarization directions. Here we exploit the analogy
with this bosonic system by specifying energy filters only. As
the spin entanglement is preserved by such filters, this opens
up the possibility for making current measurements with polarized
contacts of various orientation. Indeed, the respective polarization
orientation of a set of ferromagnetic nano-electrodes
can be controlled by an external magnetic field because
the magnetization reversal is a function of the size of the electrode.  
\begin{figure} 
\centerline{\epsfxsize=8.6cm \epsfbox{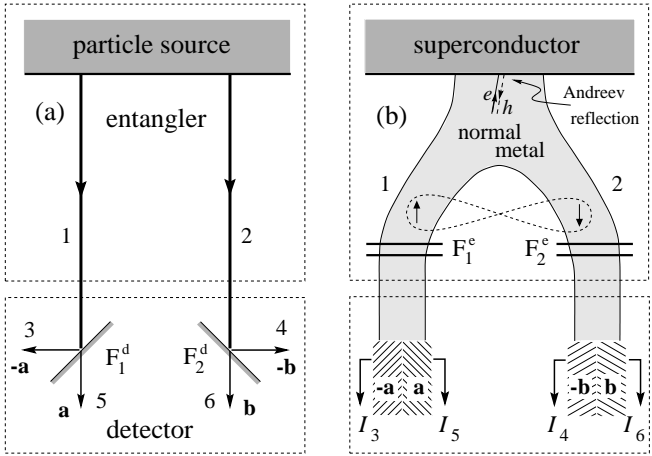}}  
\narrowtext\vspace{2mm} 
\caption{\label{fig_bell} a) Schematic setup 
for the measurement of Bell inequalities: a
source emits particles into leads 1 and 2. 
The detector measures the correlation between beams labelled
with odd and even numbers. Filters F$^{\rm d}_{1(2)}$ select
the spin: particles with polarization along the direction 
$\pm{\bf a}(\pm{\bf b})$ are transmitted through filter 
F$^{\rm d}_{1(2)}$ into lead 5 and 3 (6 and 4). b)
Solid state implementation, with
superconducting source emitting Cooper pairs into the leads. 
Filters F$^{\rm e}_{1,2}$ (Fabry-Perot
double barrier structures or quantum dots) prevent Cooper
pairs from entering a single lead. Ferromagnets with
orientations $\pm{\bf a},~\pm{\bf b}$ play the role of 
filters F$^{\rm d}_{1(2)}$ in the a): they are transparent for
electrons with spin aligned along their magnetization.}   
\end{figure}  
Another point that differs from the photon experiments is that {\it transport}
(average current, or noise correlations averaged over time) are
typically measured, while it is possible to detect photons one
by one. Nevertheless, at long times, 
noise correlations at zero frequency can be connected to number
counting correlations. Let $N_\alpha(\tau)$ denote the number
of electrons detected in a time interval $\tau$ and the
corresponding current noise
correlations spectrum $S_{\alpha\beta}(\omega)$.
In the limit of large times, the number correlator associated
with two different leads reads:
\begin{equation}
   \label{cor}
   \langle N_\alpha(\tau)N_\beta(\tau)\rangle\approx \langle
   I_\alpha\rangle \langle I_\beta\rangle \tau^2+\tau S_{\alpha\beta},
\end{equation}
Here, ``large'' times mean that 
$\omega_0^{-1}\ll \tau \ll \omega_{fl}^{-1}$, where $\omega_{fl}$ is 
the lower threshold frequency for $1/f$ noise, and
$\hbar\omega_0={\mathrm Min}(e|V|,\Gamma)$ is the upper frequency 
associated either with the voltage or with the energy width 
$\Gamma$ of the filters.
It is then possible to transcribe the Bell
inequalities expressed usually as correlators of
numbers of particles \cite{chtchelkatchev}
in terms of noise correlators. 
Care must be taken so that the contribution 
of the reducible products $\langle
N_\alpha(\tau)\rangle \langle N_\beta(\tau)\rangle$
can be safely neglected. This can be achieved 
by reducing the transparency of the S-SWNT
interface.
Here we only mention the main result of Ref.
\cite{chtchelkatchev}. 
The typical geometry for testing Bell inequalities, given a
source of particles, is depicted in Fig. \ref{fig_bell}a.
The condensed matter implementation which exploits the filters
described in the previous sections appears in 
Fig. \ref{fig_bell}b. Here it is argued that the 
measurement of time dependent currents  in the different leads of
this  device, and the subsequent computation of the cross
correlation functions, can lead to a Bell inequality test. On
general grounds,  these noise correlations
$S_{\alpha\beta}(\omega=0)$ with arbitrary polarizations in
leads $\alpha$ and $\beta$ have two contributions:
\begin{equation}    \label{S_ideal}
   S_{\alpha\beta}=
   S_{\alpha\beta}^{(a)}\sin^2\left(\frac{\theta_{\alpha\beta}}
   2\right)+
   S_{\alpha\beta}^{(p)}\cos^2\left(\frac{\theta_{\alpha\beta}}
   2\right),
\end{equation}
where, $\theta_{\alpha\beta}$ denotes the angle between the magnetization
of leads $\alpha$ and $\beta$. $S_{\alpha\beta}^{(a(p))}$
is the noise power in the special situation where the 
orientations of the two ferromagnets are {\it
antiparallel} ({\it parallel}).  
Let us assume that the energy width of the filters is given by 
The Bell inequality can be expressed in terms of these two
quantities
 \begin{equation}
   \label{Bell-maximum}
   \left|\frac{S_{\alpha\beta}^{(a)}-S_{\alpha\beta}^{(p)}}
   {S_{\alpha\beta}^{(a)}+S_{\alpha\beta}^{(p)}}\right|\leq \frac
   {1}{\sqrt 2}.
\end{equation}
However, if the only transfer process at the boundary is
Andreev reflection (no quasiparticle transmission between the
two leads) $S_{\alpha\beta}^{(p)}=0$,
so that the Bell inequality is maximally violated.
Note that the presence of $2$ propagating modes 
in each arm, and possible channel mixing, does not spoil
the detection,  because all the 
quantities are summed over channel numbers.
The result 
of Eq. (\ref{Bell-maximum}) confirms that a rigorous test for
entanglement can be reached within a scattering approach.

Order of estimates for the current and the noise correlations 
are obtained with the assumption that 
the Andreev cross-reflection probability is 
denoted $R_A$. Neglecting the angle dependence of the filters, 
and assuming that electron transmission from the superconductor,
then through one of the filter is sequential, gives the current
estimate $\langle I_\alpha \rangle\sim e R_A\Gamma/\hbar$. Using
the Schottky formula this yields: 
\begin{equation}
S_{\alpha\beta}^{(a)}\sim e^2R_A\Gamma/\hbar~.
\end{equation}
Requiring that the reducible number correlator be 
negligible compared to the irreducible correlator 
(the one connected to the noise) thus yields, together 
with $\langle N_\alpha \rangle \sim \tau \langle I_\alpha
\rangle$ and Eq. (\ref{cor}) : 
\begin{equation}
{\mathrm Max}\left( {\hbar\over \Gamma},
{\hbar\over e|V|}\right) < \tau < {\hbar\over \Gamma R_A}~.
\end{equation}
This means that strictly speaking, if a Bell test is to be
performed on this electronic subsystem, the acquisition time, or
measurement time, is bounded from above. Note that this is no
different from the situation with quantum optics \cite{aspect},
in which photons are detected by coincidence counting. In order 
to probe entanglement, one needs to distinguish between 
the two photons of an entangled pair generated by parametric 
down conversion, and two photons which belong to two 
distinct pairs, which are uncorrelated.  
In our solid state setting, this would limit our
approach to poorly transmitting Andreev interfaces, or
alternatively to extremely selective energy filters. 
Note however that once the assumption is satisfied, the final
expression which is to be checked, Eq. (\ref{Bell-maximum}), 
contains only zero frequency noise correlators, and is 
independent on this acquisition time.  

The cross--correlation geometry depicted in Fig.
\ref{fig_bell} may be difficult to implement in experiments,
as ideally electrons with both spin orientations 
$\pm{\bf a}(\pm{\bf b})$ are to be collected in
the ``double'' leads. Other experimental 
geometries with two leads, based
on the violation of Clauser--Horne inequalities \cite{clauser} (a
variant of Bell inequalities) can possibly be implemented in a
more straightforward manner as each lead is attached to a
{\it single}  ferromagnet.

\section{Single electron picture vs. Luttinger liquid picture}

We now enquire how the transport properties of the 
S-SWNT device can be modified if the nanotube is 
considered to be a strongly correlated one 
dimensional system. Close to their Fermi level,
metallic (armchair) nanotubes have an energy spectrum which 
can be approximated by two ``crosses'', corresponding 
to two one--dimensional modes. In the presence of 
Coulomb interactions, from a theoretical point of view 
the system can be considered as two Luttinger liquids
with total (relative) charge (spin) degrees of freedom
\cite{egger_review}. Interaction parameters 
$K_{j\delta}$ ($j\delta=c+,c-,s+,s-$) characterize 
the strength of the Coulomb interaction in these sectors: 
for time reversal 
symmetric situations, the spin interaction 
parameters $K_{s\delta}=1$ ($\delta=\pm$),
while in the absence of Coulomb interactions,
$K_{c+}=1$. Repulsive electron interactions 
correspond to $K_{c+}<1$.
Except for tunneling density of 
states measurements, there exist little data pointing
out to Luttinger liquid behavior when charge
propagates along the nanotube. Preliminary  
two-terminal transport measurements on suspended 
nanotubes with embedded contacts have recently
been performed \cite{roche}. According to this 
work, the low shot noise level cannot be fully understood 
with a single electron picture. 
Here, there are two fundamental aspects of Luttinger 
liquid physics which could influence transport
in the S-SWNT device.

First, consider the propagation of charge along the nanotube.
Does it remain a ``good'' wave guide in the presence of interactions ?  
According to tunneling density of states measurements
\cite{mceuen}, the Luttinger interaction parameter is ``strong''
($K_{c+}\approx 0.3$). It is then predicted \cite{kane_fisher} that 
at zero temperature, even the presence
of weak impurity scattering can lead to insulating 
behavior -- as shown by renormalization group
arguments. Thus the presence of the slightest 
concentration of impurities could impede 
electron propagation. Fortunately, at finite 
temperatures the effective impurity barrier strength 
$\lambda$ is reduced: 
$\lambda_{eff}\sim \lambda T^{K_{c+}-1}$
($\lambda$ is the bare impurity barrier strength). 
At ``high'' enough temperatures, one then
recovers a linear current--voltage behavior
with a conductance $G(T)-2K_{c+}e^2/h\sim \lambda^2 
T^{2K_{c+}-2}$. Note that the higher the temperature, 
the smaller is the deviation from ideal transmission. 
Except for renormalization of the 
free conductance by interactions,
one should then expect the single electron 
picture to hold. 

The role of Coulomb interactions can also be minimized 
if the nanotube is placed on top of a metallic or a 
doped semiconductor substrate, or when it is located
close to a metallic gate. The long range Coulomb interaction 
in the nanotube is then efficiently screened, 
increasing the interaction parameter $K_{c+}$ 
close to its non-interacting value (a thin oxide layer should be planned 
in order to avoid spurious contacts).

Next consider the injection of electrons in
the nanotube. The addition of a single
electron does not represent an eigenstate 
of the nanotube : this is explicit 
in the vanishing of the density of states at the Fermi level. 
An electron is decomposed into pairs of right and left moving 
chiral excitations with fractional charges \cite{pham_imura} 
associated with each sector (charge/spin and total/relative).
Such charges 
$Q_{j\delta}^\pm = (1\pm K_{j\delta})/2$
can in principle be detected via the combination of 
an autocorrelation noise
measurement and of a noise cross-correlation measurement
\cite{crepieux}. Such pairs of charges moving in opposite 
directions then have entangled degrees of freedom.

Next, consider the injection of two electrons 
in the nanotube.
For the S-SWNT device without filters, two electrons are 
expected to break up into two pairs of entangled 
chiral quasiparticles, which are themselves 
entangled because they originate from the same Cooper 
pair. Nevertheless, in the presence of Fermi liquid contacts
\cite{safi_maslov}, one expect to recover the essential 
features of a single electron system which is only 
correlated by the superconductor only, due to the multiple 
reflections of the quasiparticles at the contacts.

In the presence of selective energy filters or coherent 
quantum dots, which select positive and negative energies
as measured from the superconductor chemical potential, 
the situation will at first be similar. Quantum dots can 
accomodate electrons only. Quasiparticles excitations 
generated by superconductor injection will recombine 
into electrons when reaching the dots.
However, further propagation
{\it past} the dots, along the nanotube extremities
will happen once again in a 1D correlated electron system.   
It is well known that Luttinger liquids exhibit 
the phenomenon of charge--spin separation. 
Charge excitations do not propagate with the same 
velocity as spin excitations.  
Experimental claims for the observation of such 
separation of charge and spin degrees of freedom
have been made recently with semiconductor quantum 
wires obtained with cleaved edge overgrowth techniques
\cite{yacoby}. Provided that no impurities bother the 
propagation of these excitations it is likely to operate 
also in the S-SWNT device. If one places ferromagnetic 
filters further down the extremities, there will then be 
a time delay $L(v_{charge}^{-1}-v_{spin}^{-1})$ 
between the detection of a charge excitation and the 
detection of a spin excitations. While these effects 
could affect the Bell analysis, they can be 
minimized by reducing the propagation length $L$ between 
the dots and the ferromagnetic filters.  

\section{Possible Fabrication processes}
\label{fabrication_section}

We are aware that the S-SWNT device together with its {\it in-situ} 
integrated entanglement detector 
is rather difficult to fabricate. However tremendous progresses 
have been made in the last few years to make
the electrical contact between a SWNTs and a metallic 
electrode reliable and reproducible. 
Indeed, the contact resistance can be reproducibly controlled at values
 close to ideal threshold of $4e^2\over h $
\cite{prl_kasumov,nature_nt_fabry_perot}. The observation
of superconductivity induced by proximity effect in carbon
nanotubes \cite{proximity supra-NT} confirms that correlated 
electrons can effectively propagate over mesoscopic lengths in
SWNTs. We present here several possible directions that
could be followed to succeed in realizing such experiment.

First like in most reported experiments for which a metal-SWNT-metal junction
 could be achieved with a resistance lying
 in the 10 {\it k}$\Omega$ range, the SWNT position could be localized by scanning probe
 or by transmission electron microscopy.
Contacts could be then performed in a second step either
 by depositing noble metal on top \cite{defect_NT,SETintraNT,Bachtold_NTlogic,avouris_device}  
or by bonding the SWNT with a laser pulse \cite{proximity supra-NT}. 
 It will then involve multilevel deposition of metallic thin films,
 using aligned masks made by electron beam lithography.

Other fabrication schemes could be envisioned even if first published results 
have concluded to an increased contact resistance with respect to the previously described method:
 they involve the fabrication of the metallic contacts as the first step with a rather clean top surface.
One could then proceed to the SWNT deposition. 
There are three different possibilities for that last step : 1) deposition of ex-situ
synthesized SWNT, followed by AFM manipulation \cite{Postma,manipul_NT_samuelson,SET_ROSCHIER}, 2) deposition of
ex-situ synthesized SWNT with self assembly guided by functionalized 
electrodes \cite{bourgoin} 
3) direct in-situ  growth of SWNT from contacting catalytic electrodes \cite{marty}.
These two latter methods allow batch processing and
avoid any alignment step.  Many sample could be performed in
parallel which provide a great advantage with respect to all
other fabrication schemes that involves time-consuming alignment
steps. Ferromagnet pads and/or control gates could be deposited
before or after the SWNT contacting step depending on  the
process.   In the light of the experimental achievements of Refs.
\cite{Bezryadin,SETintraNT}, it is tempting to make a specific suggestion
-- however realistic -- for the Bell inequalities device of section
\ref{bell_subsection}.

%
%
\begin{figure} 
\centerline{\epsfxsize=6.0cm \epsfbox{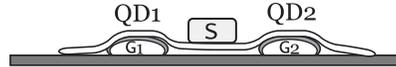}}  
\narrowtext\vspace{4mm} 
\caption{Nanotube coupled to three metallic electrodes : the
central electrode is superconducting and constitutes a good
contact with the nanotube, while the two other electrode stand below the tube : 
the former act as a source of correlated electrons while the latter
 act as voltage gates and induce kinks in the tube creating local barriers 
that define the energy filters of section V}  
\end{figure}  
Pursuing the suggestion of section \ref{splitter_section}, 
assume that a SWNT has been deposited in good contact with  a
superconducting island. Now, on each side  of the superconductor
(Fig. 7), two normal metal islands have also
been deposited. Because of the bending of the nanotubes and the
resulting barriers due to bending defects, the latter islands
allow to define two quantum dots [22,53] which play the role of energy
filters. These islands are capacitively coupled to gate electrodes allowing
fine tuning of the resonant level in each dot.
The resulting device constitutes the Cooper pair splitter with
filters. Both ends of the nanotube could therefore be
contacted to ferromagnetic leads with varying orientation 
in order to provide a Bell type test.  Finally, note that
coincidence measurements for single
electron events are likely to be feasible in the near future due
to recent improvements in the instrumentation of single electron
transistors \cite{rf_SET}. An increase of the 
detection  bandwidth of single electron events is still
required. 

\section{Conclusion}

In conclusion, we have proposed a single nanotube which plays
the role of a normal superconducting beam splitter.
This splitter can be exploited to detect positive (bosonic)
correlations in {\it a priori} purely Fermionic system. 
In this first step either metallic of semiconducting
nanotubes (with a back-gate) could be used.
Moreover, in relation
to the ongoing interest in quantum information processing, the
addition of filters which select either electron spin or energy
could provide a robust scheme for generating
entangled pairs electrons at the boundary of a superconductor. 
This second experiment is likely to be successful using
semiconductor nanotubes because excessive screening
in metallic nanotubes would render the gates less efficient. 
This nanotube device combines state of the art technology in
both metal/superconductor lithography and manipulation/growth of carbon
nanotubes. Complications associated with the correlated nature 
of the electron state in the nanotube -- the Luttinger liquid --
have been addressed 
qualitatively, and indicate that the working temperature should be 
chosen large enough to minimize the effect of impurities, 
but low enough to preserve quantum coherence in the device. 

On one hand, it would allow for the first time to
perform an EPR experiment on massive particles with fermionic
statistics. On the other hand, it could become an useful
device for quantum information processing.  \vspace{-0.3truecm}

\acknowledgements

G.B.L. and N.C acknowledge support from: the Russian 
science-support foundation, Russian foundation for basic
research (grant N 000216617), Russian ministry of science
("Quantum macrophysics" project), Swiss National Foundation,
Netherland foundation for collaboration with Russia.
N.C. acknowledges support from agreements between Landau Institute and Ecole
Normale Sup\'erieure de Paris.

\end{multicols}
\end{document}